\documentclass{iaus}
\usepackage{graphicx}

\title[Images of B\[e\] stars]
      {
        Images of unclassified and supergiant B[e] stars disks with
        interferometry
      }
      
      \author[F. Millour et al.]   
             {
               F. Millour$^{1,2}$, A. Meilland$^1$, O. Chesneau$^2$,
               M. Borges Fernandes$^{3}$, J. H. Groh$^1$, T. Driebe$^{1,4}$,
               A. Liermann$^1$, \& G. Weigelt$^1$
             }
             
             \affiliation{$^1$Max-Planck Institute for Radioastronomy, auf dem
               H\"ugel 69, 53121 Bonn, Germany
               \\[\affilskip]
               $^2$Observatoire de la c\^ote d'Azur, Bd. de
               l'Observatoire, 06304 Nice, France
               \\[\affilskip]
               $^3$Observatorio Nacional, Rua General Jos\'e Cristino,
               77, Rio de Janeiro, Brazil
               \\[\affilskip]
               $^4$German Aerospace Center (DLR), K\"onigswinterer
               Str. 522-524, 53227 Bonn, Germany
               \\ Contact email: {\tt fmillour@oca.eu}}
             
             \pubyear{2010}
             \volume{xxx}  
             \pagerange{119--126}
             \jname{Active OB stars: Structure, Evolution, Mass loss
               and critical limits}
             \editors{C. Neinert, B.D. Editor \& C.E. Editor, eds.}
             
\begin{document}

             \maketitle

             \begin{abstract}
               B[e] stars are among the most peculiar objects in the
               sky. This spectral type, characterised by allowed and
               forbidden emission lines, and a large infrared excess,
               does not represent an homogenous class of objects, but
               instead, a mix of stellar bodies seen in all
               evolutionary status. Among them, one can find Herbig
               stars, planetary nebulae central stars, interacting
               binaries, supermassive stars, and even ``unclassified''
               B[e] stars: systems sharing properties of several of
               the above. Interferometry, by resolving the innermost
               regions of these stellar systems, enables us to reveal
               the true nature of these peculiar stars among the
               peculiar B[e] stars.
               \keywords{
                 techniques: high angular resolution --
                 techniques: interferometric --
                 binaries: close --
                 stars: individual HD87643, HD62623 --
                 stars: mass loss --
                 stars: winds, outflows
}
             \end{abstract}

             \firstsection 

             \section{Introduction}

             We started an observing program covering the brightest
             unclassified and (candidate) supergiant B[e] stars, by
             using the Very Large Telescope Interferometer. For now,
             about 10 targets have been observed using a combination
             of AMBER (near-IR) and MIDI (mid-IR) observations. Here
             we focus on imaging and therefore stick to AMBER
             observations of two of the targets: the unclassified B[e]
             star HD87643 and the supergiant A[e] star HD\,62623.

             \section{HD\,87643}
             
             We used an extensive dataset spanning several orders of
             magnitude of spatial resolution to partly unveil the
             nature of this stunning system.

             {\bf Wide-field images:} Many new details are seen in our
             wide field image of the nebula around HD87643: “blown up”
             structures in the large scale nebula, which could be the
             result of an eruption much like in LBVs, and a series of
             arcs, unseen before, which could be the result of
             periodic ejections of matter.

             {\bf Interferometric images:} Our interferometric images,
             made using the AMBER/VLTI instrument, show a companion
             star to the bright star. In addition, the primary star
             exhibits an extended shell (~4\,mas), and background
             emission is detected.
             
	     {\bf Interpretation:} Using a model involving both
             stellar components plus a resolved background, we were
             able to separate individual spectra in H, K and N
             bands. We find that the primary star is enshrouded in a
             dust shell heated at the dust-sublimation temperature: we
             resolved the inner region of the HD87643 disk. The
             secondary component is also enshrouded in dust, at
             temperatures ranging from 500 to 50K. The resolved
             background holds most of the silicate emission, hence
             being likely composed of colder condensed dust. As a
             conclusion, and since the distance to the system is still
             poorly constrained, HD87643 could be a YSO instead of an
             evolved object. See \cite{Millour2} for more details.

             \begin{figure}[htbp]
               \begin{center}
                 \includegraphics[width=0.95\textwidth,
                   angle=-00]{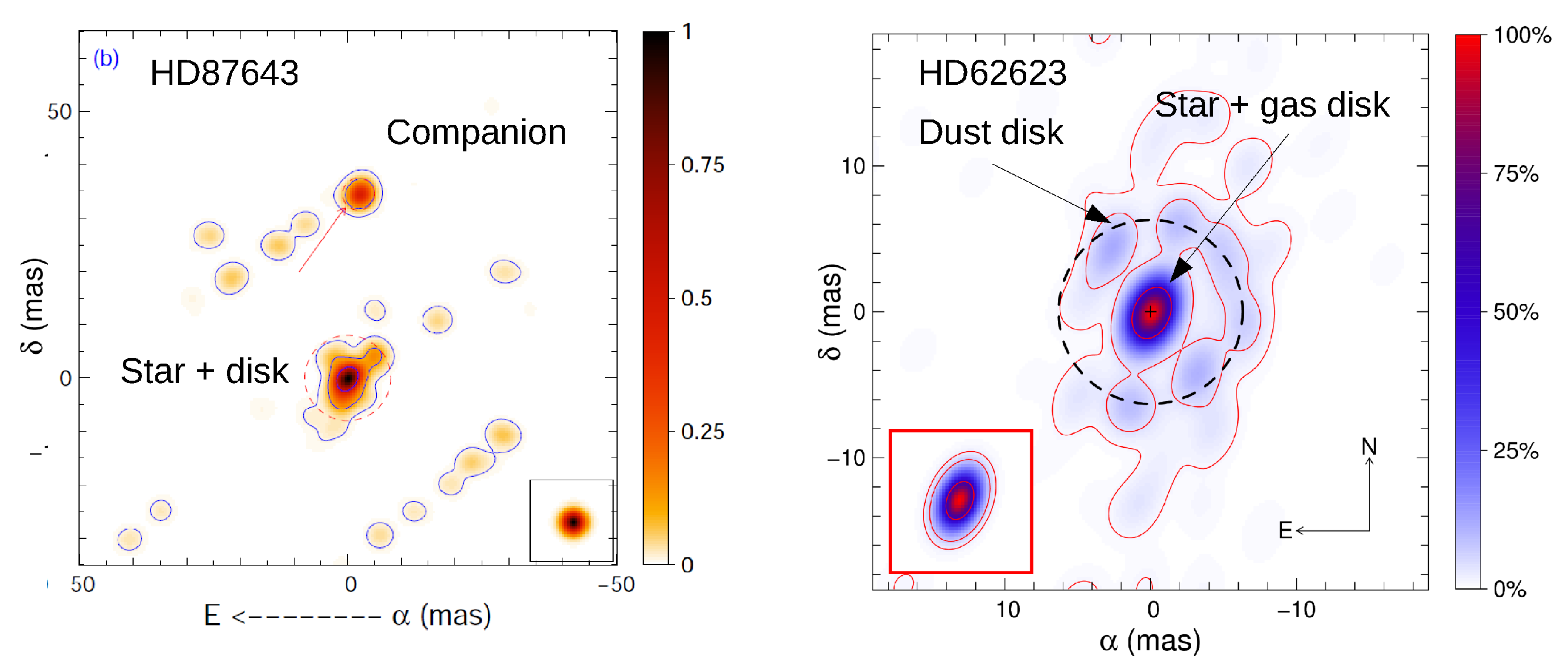}
                 \caption{
                   Left: AMBER/VLTI image of the star HD87643,
                   revealing a companion star and a disk around the
                   primary star. Right: AMBER/VLTI image of the
                   previously-putative disk around the star HD62623. A
                   spectrally-resolved image-cube could be
                   synthesised, evidencing Keplerien rotation in this
                   disk.
                 }
                 \label{fig1}
               \end{center}
             \end{figure}

             \section{HD\,62623}

             This star was already extensively studied by spectroscopy
             \cite{Pletsetal}. Hence it was already known that it
             hosted a circumstellar disk.
             
             {\bf The continuum:} We clearly resolve the gas and the
             dust disk in the continuum, imaging for the first time
             the disk of an evolved star. The image shows an outer
             ring corresponding to the inner rim where dust
             sublimation occurs, while we also resolve a central
             region where free-free emission from the gas takes
             place. The dust emission is asymmetric, which we
             attribute to an inclination effect. This direct picture
             strikingly confirms what was indirectly inferred from
             spectroscopic data previously.

             {\bf The Br$\gamma$ line:} Our high-spectral resolution
             image cube in the Br$\gamma$ line directly shows the
             motion of the gas in different velocity bins. The
             combination of a model of the dust, which gives the
             on-sky orientation and inclination of the disk, and of a
             model of the rotating gas, allows us to claim that we
             detected Keplerian rotation in that disk. We can now
             clearly rule out expanding wind models for
             HD\,62623. Instead, we have a Keplerian rotating disk,
             where expansion is negligible. Such a disk, unexpected
             for massive stars, is common in young stellar objects
             like Herbig stars. In the case of HD\,62623, the presence
             of a very close unseen companion, previously detected by
             radial velocities, might be the key to the formation of
             such a rotating disk.

             \end{document}